\documentstyle[prd,aps,eqsecnum,preprint,tighten]{revtex}
\draft
\begin{document}
\title{Regularizing Property of the Maximal Acceleration Principle in Quantum
Field Theory}
\author{V.V. Nesterenko\thanks{E-mail: nestr@thsun1.jinr.ru}}
\address{Bogoliubov Laboratory of Theoretical Physics,
JINR, Dubna, 141980, Russia}
\author{A. Feoli$^{b,c}$,
G. Lambiase$^{a,b}$\thanks{E-mail:
lambiase@physics.unisa.it}, G.~Scarpetta$^{a,b}$}
\address{ $^a$Dipartimento  di  Scienze Fisiche ``E.R. Caianiello'',
Universit\`a di Salerno, Italia.\\
$^b$Istituto Nazionale di Fisica Nucleare, Sez. di Napoli.\\
$^c$Facolt\`a d'Ingegneria, Universit\`a del Sannio, Benevento, Italia
}
\date{\today}
\maketitle
\begin{abstract}
It is shown that the introduction of an upper limit to
the proper acceleration of a particle can smooth the problem
of ultraviolet divergencies in local
quantum field theory. For this aim, the classical model of a relativistic
 particle with maximal proper acceleration is quantized canonically
by making use of the generalized
Hamiltonian formalism developed by Dirac. The equations for the wave function
are treated
as the dynamical equations for the corresponding quantum field. Using the Green's function
connected to these wave equations as  propagators in the Feynman integrals leads to an
essential improvement of their convergence properties.
\end{abstract}
\thispagestyle{empty} 

\section{Introduction}

During the last years,  strong evidences have arosed
 in different areas of theoretical physics, that the proper acceleration of
 elementary particles (in general case, of
any physical object) cannot be arbitrary large, but it should be superiorly limited
by some universal value ${\cal A}_m$\footnote{We are using the unit system where
$\hbar =c=1$.
Therefore the dimensions of acceleration and mass are the same.}
(maximal proper acceleration).
For instance, in string theory \cite{GSW,BN}, seeking to present a unified
description of all fundamental interactions, including gravity,
it was derived that  string
acceleration
must be less than some critical value, determined by the string tension
and its mass~\cite{SV,G,FS,Feoli,MG}. Otherwise in string dynamics the Jeans--like
instabilities arise,  which lead to unlimited grow of the string length \cite{SV,G,FS}.

On the other hand the theories of the fundamental and hadronic strings unambiguously
predict an upper limit to the temperature in the thermodynamical ensemble of the
strings. It is due to an extremely fast (exponential) grow of the number of levels
 in the string
spectrum when the energy or mass raise~\cite{BN,HW}. At the critical temperature, the
statistical weight of the energy levels completely suppresses the Boltzmann factor
$\exp{(-E_n/T)}$ and, as a result, the statistical sum of the string ensemble proves to
be infinite. In hadronic physics this critical temperature is  the
Hagedorn temperature\footnote{R. Hagedorn~\cite{Hag}
 derived this critical temperature in the framework of
bootstrap description of the hadron dynamics,  before the development of the
hadronic string models.}. Further, there is a well--known relation between
 the acceleration of
an observer, $a$, and the temperature of photons bath  detected,
 $a=2\pi T$
(Unruh's effect~\cite{Unruh}). Thus the Hagedorn critical temperature gives
 rise  to an upper
limiting value of the proper acceleration.

In the framework of an absolutely different approach, a conjecture about the
existence of a maximal proper acceleration of an elementary particle was
introduced by E.R. Caianiello~\cite{Caia}.
In these papers a new geometric setting for the quantum mechanics has been
developed in which the quantization was
interpreted as introduction of a curvature in the relativistic
 eight dimensional space--time tangent bundle
$TM = M_{4}\otimes TM_{4}$, that
incorporates both the space--time manifold $M_4$ and the momentum space $TM_4$.
The standard operators of the Heisenberg algebra, $\hat{q}$ and $\hat{p}$ are
represented as the covariant derivatives in $TM$, the quantum commutation relations
being treated as the components of the curvature tensor. It is remarkable,
that the line element in $TM_8$
intrinsically involves an upper limit on the proper acceleration of the particle
\cite{CGPS,IJTP}.

The existence of the upper bound on the proper acceleration is intrinsically connected
with the extended nature of the particle or string. Therefore one can
expect that quantum field theory, involving maximal proper
 acceleration, could be free of  ultraviolet divergencies,
originated by the point--like character of the particles in
local quantum field theory, or, at least,that the degree of
these divergencies could be lower.

Without claiming to solve the problem of constructing a new quantum field theory
inglobing
the principle of maximal acceleration, this note seeks to present some convincing
evidences that, in fact, the bond on the maximal acceleration can at least smooth
the problem of divergencies. The starting point of our consideration will be the
classical model of a relativistic particle with maximal proper
acceleration~\cite{CGPS,IJTP,NFS}. Upon quantizing this model the equations for the wave
functions can be treat as the dynamical equations for the corresponding field function.
Finally we are interested to the Green's function for this new field equations,
specifically, to the behaviour of this function in momentum space.
It will be shown that  using this Green's function as the propagator in the Feynman
integrals leads   to an essential  improvement of the convergence properties of the
latter.

The layout of this paper is as follows. In Sect.\ II the classical dynamics of the
relativistic particle with maximal proper acceleration is presented both in the
Lagrangian and in the Hamiltonian forms. Sect.\ III is devoted to the quantum theory of
this model. In Sect.\ IV (Conclusion) the arguments presented in favor of the
regularizing role of the maximal proper acceleration are shortly discussed.

\section{Relativistic Particle with Maximal Proper Acceleration}
Let $x^{\mu}(s),\; \mu=0,1,\ldots , D-1$ be the world trajectory of a particle in the
$D$ dimensional Minkowski space-time with the Lorentz signature $(+,-,\ldots ,-)$. We
are using here the natural parameterization of the particle trajectory,
$ds^2=dx_{\mu}dx^{\mu}$. From the geometrical point of view the proper acceleration of
the particle is nothing else as the curvature of its trajectory~\cite{Eisenhart}
\begin{equation}\label{curvature} k^2(s)=-\frac{d^2x_{\mu}}{ds^2}\frac{d^2x^{\mu}}{ds^2}\,{.} \end{equation}
In a complete analogy with the action for an usual spinless relativistic particle
\begin{equation}\label{action} S_0=-m\int ds \end{equation}
the action of a particle with upper bonded
acceleration is given by the formula \cite{CGPS,IJTP,NFS}
\begin{equation}\label{action1}
S=-\mu_0\int \sqrt{{\cal A}_m^2-k^2(s)}\,ds\,{.}
\end{equation}
Here $\mu_0=m/{\cal A}_m$ and ${\cal A}_m$ is the
maximal proper acceleration of the particle. When ${\cal A}_m\to \infty$ the action
(\ref{action1}) reduces to (\ref{action}).

In paper \cite{NFS} the classical dynamics generated by the action (\ref{action1}) has
been investigated completely in the framework of a new method for integrating the
equations of motion for the Lagrangians with arbitrary dependence on the particle proper
acceleration. It was shown, in particularly, that the
particle acceleration in this model always obeys the condition $k^2(s)<{\cal A}_m^2$.

In order to  quantize this model, one has to develop the Hamiltonian
formalism. For this purpose an arbitrary parameterization of the particle trajectory
$x^{\mu}(\tau)$ should be considered, where the evolution parameter $\tau$ is subjected
only to the condition $\dot{x}^2>0$. Dot means differentiation with respect $\tau$. In
the $\tau$--parameterization the proper acceleration of the particle is determined by
the formula \begin{equation}\label{curvature1}
k^2(\tau)=\frac{(\dot{x}\ddot{x})^2-\dot{x}^2\ddot{x}^2}{(\dot{x^2})^3}\,{.}
 \end{equation}
 When
passing from the natural parameterization $(s)$ to the arbitrary evolution parameter
$(\tau)$, the relations~\cite{NFS} \begin{equation}\label{relations}
\frac{d}{d\tau}=\sqrt{\dot{x}^2}\frac{d}{ds}\,{,} \quad \frac{d^2x_{\mu}}{ds^2}=
\frac{\dot{x}^2\ddot{x}_{\mu}-(\dot{x}\ddot{x})\dot{x}_{\mu}}{(\dot{x}^2)^2}\,{,} \quad
k\frac{\partial k}{\partial \ddot{x}_{\mu}}=-\frac{1}{\dot{x}^2}
\frac{d^2x_{\mu}}{ds^2}\,{,} \end{equation} $$ k\frac{\partial k}{\partial
\dot{x}_{\mu}}=\frac{1}{(\dot{x}^2)^4}\,\{
\dot{x}^2(\dot{x}\ddot{x})\ddot{x}^{\mu}+[2\dot{x}^2\ddot{x}^2-3(\dot{x}\ddot{x})^2]
\dot{x}^{\mu}\} $$ prove to be useful. According to Ostrogradskii~\cite{W,N1}, the
Hamiltonian description of the model in question requires introduction of the following
canonical variables \begin{equation}\label{canvar} q_{1\mu}=x_{\mu}, \quad
q_{2\mu}=\dot{x}_{\mu}\,{,} \end{equation} $$ p_1^{\mu}=-\frac{\partial L}{\partial
\dot{x}_{\mu}}-\frac{dp_2^{\mu}}{d\tau}\,{,}\quad p_2^{\mu}=-\frac{\partial L}{\partial
\ddot{x}_{\mu}}\,{,} $$ where $L$ is the Lagrange function in the
$\tau$--parameterization.

The invariance of the action (\ref{action1}) under the transformations
$x^{\mu}\to x^{\mu}+a^{\mu}$, $a^{\mu}=\text{const}$ entails the
conservation of the
energy--momentum vector $p_1^{\mu}$. Hence in our
consideration the mass should be defined by
\begin{equation}\label{mass}
M^2=p_1^2\,{.}
\end{equation}
In contrast to the usual relativistic particle with the action (\ref{action})
the mass of the particle with restricted  proper acceleration, in general case, does not
coincide with the parameter $m$ entered its action (\ref{action1}).

The invariance of the action (\ref{action1}) under the Lorentz transformations
leads to conservation of the angular momentum tensor
\begin{equation}\label{angmom}
M_{\mu\nu}=\sum_{a=1}^{2}(q_{a\mu}p_{a\nu}-q_{a\nu}p_{a\mu})\,{.}
\end{equation}
Usually the tensor $M_{\mu\nu}$ is used for constructing the spin variable $S$.
In the case of the $D$--dimensional space--time $S$ is defined
by~\cite{Schweber}
\begin{equation}\label{spin}
S^2=\frac{W}{M^2}\,{,}
\end{equation}
where
$$
W=\frac{1}{2} M_{\mu\nu}M^{\mu\nu}p_1^2-(M_{\mu\sigma}p_1^{\mu})^2\,{,}\quad M^2=
p_1^2\,{.}
$$
For $D=4$ the invariant $-W$ is the squared Pauli--Lubanski vector
\begin{equation}\label{pauli}
W=-w_{\mu}w^{\mu}, \qquad w_{\mu}=\frac{1}{2}\varepsilon_{\mu\nu\rho\sigma}
M^{\nu\rho}p_1^{\sigma}\,{.}
\end{equation}
Obviously spin $S$ is also a conserved quantity.

An essential distinction of the  Lagrangians with higher derivatives, like
(\ref{action1}), is the following~\cite{NFS}. The mass $M$ and the spin $S$ are not
expressed in terms of the parameters of the Lagrangian, but are  integrals of
motion, whose specific values  should be determined by the initial conditions for the
equations of motion~\footnote{The spin $S$
of the usual relativistic particle with the action (\ref{action}), calculated by the
formulae (\ref{spin}) or (\ref{pauli}), identically equals zero because in this case there is only one pair
of the canonical variables $q,\,p$.}. This point considerably complicates
transition to the secondary
quantized theory (quantum field theory) starting just from the action (\ref{action1}).
Usually the wave equation for quantum field uniquely specifies the mass and the spin of
the particles described by this field \cite{BSh}.

The action (\ref{action1}), as well as (\ref{action}), is invariant under the
reparameterization $\tau\to f(\tau )$ with an arbitrary function $f(\tau )$. As a result
there should be the constraints in the phase space \cite{BNF}. Using the definition of
$p_2$ in (\ref{canvar}) and formulae (\ref{relations}) one easily deduces the primary
constraint
\begin{equation}\label{constrain1} \phi (q,p)=p_2^{\mu}q_{2\mu}\approx 0\,{,}
\end{equation}
where
$\approx$ means a weak equality \cite{Dirac}. There are no other primary constraints in
the model under consideration because the rank of the Hessian matrix $$ \frac{\partial^2
L}{\partial\ddot{x}_{\mu}\partial\ddot{x}_{\nu}} $$ equals $D-1$. The canonical
Hamiltonian \begin{equation}\label{Hamiltonian}
H=-p_1^{\mu}\dot{x}_{\mu}-p_2^{\mu}\ddot{x}_{\mu}-L\,{,} \end{equation} rewritten in terms of the
canonical variables $q_a, p_a,\;\;
 a=1,2,$ becomes
\begin{equation}\label{Hamiltonian1}
H=-p_1q_2+{\cal A}_m\sqrt{q_2^2(\mu_0^2-q_2^2p_2^2)}\,{.}
\end{equation}

The equations of motion in the phase space
\begin{equation}\label{eqmotion}
\frac{df}{dt}=\frac{\partial f}{\partial t}+\{f, H_{\text T}\}
\end{equation}
are generated by the total Hamiltonian
\begin{equation}\label{genH}
H_{\text T}=H+\lambda (\tau )\phi (q, p)\,{.}
\end{equation}
Here $f$ is an arbitrary function of the canonical variables and of the evolution
parameter $\tau$, $\lambda (\tau)$ is the Lagrange multiplier. The Poisson
brackets are defined in a standard way
\begin{equation}\label{Poisson}
\{f, g\}=\sum_{a=1}^{2}\left(\frac{\partial f}{\partial p_a^{\mu}}
\frac{\partial g}{\partial q_{a\mu}}-\frac{\partial f}{\partial q_a^{\mu}}
\frac{\partial g}{\partial p_{a\mu}}\right)\,{.}
\end{equation}

Requirement of the stationarity of the primary constraint (\ref{constrain1})
\begin{equation}\label{phitau} \frac{d\phi}{d\tau}=\{\phi, H_T\}\approx 0 \end{equation} leads to the
secondary constraint \begin{equation}\label{constrain2} \{\phi, H_T\}=\{\phi, H\}=H\approx 0\,{.}
\end{equation} As one could anticipate, the canonical Hamiltonian vanishes in a weak sense. It is
a direct consequence of the reparameterization invariance of the initial action
(\ref{action1}) \cite{Dirac}.
Obviously, at this step the procedure of the constraint
generation terminates. Thus in this model  there are only two constraints
(\ref{constrain1}) and (\ref{constrain2}), and they are of the first class because
\begin{equation}\label{class1} \{\phi, H\}\approx 0\,{.} \end{equation}

For the spin variable $S$ we deduce
 from Eqs. (\ref{spin}) or
(\ref{pauli})
\begin{eqnarray}
\label{spin1} S^2p_1^2&=&p_1^2p_2^2q_2^2+2(p_1p_2)(p_1q_2)(p_2q_2) \\ &&
-(p_1p_2)^2q_2^2- p_1^2(p_2q_2)^2-p_2^2(p_1q_2)^2\,{.} \nonumber
\end{eqnarray}
The energy--momentum vector squared $p_1^2$ and the spin squared
$S^2$ are the integrals of motion \cite{NFS}. In the Hamiltonian dynamics
this implies
\begin{equation}\label{motionp1}
\frac{dp_1^2}{d\tau}=\{p_1^2, H_{\text T}\}\approx 0\,{,}
\end{equation}
\begin{equation}\label{motionS}
\frac{d S^2}{d\tau}=\{S^2, H_{\text T}\}\approx 0\,{.}
\end{equation}
Specifying the values of $p_1^2$ and $S^2$
\begin{eqnarray}
\label{valuep1}
p_1^2&=&M^2\,{,} \\
\label{valueS}
S^2(p,q)&=&S^2
\end{eqnarray}
we pick  a certain sector in the classical dynamics of the model,
  which is invariant with
respect to the evolution in time.

Further, it is very convenient to use the natural parameterization of the particle
trajectory putting \begin{equation}\label{onshell} \dot{x}^2=q_2^2=m^{-2}\,{.} \end{equation}
 It is remarkable that this condition is in accordance
with the Hamiltonian equations of motion \begin{equation}\label{constrain3} \{q_2^2, H_{\text
T}\}\approx 0\,{,} \end{equation} the Lagrange multiplier $\lambda (\tau )$ being undetermined.
Therefore Eq.\ (\ref{onshell}), on the same footing as Eqs.\ (\ref{valuep1}) and
(\ref{valueS}), should be treated as an invariant relation rather than a gauge
condition \cite{LCA}. In the parameterization (\ref{onshell}) the canonical momenta
$p_1$ and $p_2$ prove to be proportional to $\dot{x}=q_2$ and $\ddot{x}$, respectively.
Hence, we have now \begin{equation}\label{constrain4} p_1p_2\approx 0 \end{equation} in view of
(\ref{constrain1}) or (\ref{onshell}).

\section{Quantum theory}

Transition to quantum theory is accomplished in a standard way by imposing the
commutators \begin{equation}\label{commutation} [\hat A, \hat B]=i\{A, B\}\,{,} \end{equation} where
$\{\ldots,\, \ldots\}$ are the Poisson brackets (\ref{Poisson}), $A$ and $B$ are
arbitrary functions of the canonical variables $q_a,p_a,\;\; a=1,2$. All the
constraints in the model concerned  are of first--class  and the gauge is not fixed,
thus there is no need to
use the Dirac brackets in our consideration~\cite{Dirac}.

In quantum theory we are interested in Eqs.\ (\ref{valuep1}) and (\ref{valueS}) which
should be imposed as the conditions on the physical state vectors $\vert\psi>$
\begin{equation}\label{statep1} (p_1^2-M^2)\vert\psi>=0\,{,} \end{equation} \begin{equation}\label{stateS}
S^2\vert\psi>=S(S+D-1)\vert\psi>=0\,{,} \quad S=0,1,2, \ldots \,{,} \end{equation} where $D$ is
the dimension of space--time. The classical value of the spin variable $S^2$
has been
substituted here by the eigenvalues of the spin operator $S^2$ in  the $D$-dimensional
space-time, $S(S+D-1)$. These equations fix the values of the Casimir operators of the
Poincare group specifying the irreducible
 representation
of this group in terms of the state vectors $|\psi >$
 and providing in this way the basis for consistent description of
the elementary particles in the relativistic quantum theory.

As was noted in Introduction we shall treat equations (\ref{statep1}) and (\ref{stateS})
in two-fold way, as the equations for the wave function $|\psi >$ in the quantum
mechanics of the model in hand and as the wave equations for the corresponding
relativistic quantum field $\Phi (q_1,q_2)$ (the secondary quantized theory). This
field, as well as the wavefunction $\vert\psi>$, should depend on two arguments
$q_1^{\mu}=x^{\mu}$ and $q_2^{\mu}=\dot{x}^{\mu}$ which are treated as the canonical
coordinates in the Ostrogradskii formalism \cite{W,N1}.

Before proceeding further we have to take into account the constraints. Substituting
Eqs.\ (\ref{constrain1}), (\ref{constrain2}), (\ref{onshell}) and (\ref{constrain4})
into Eq. (\ref{spin1}) we obtain
\begin{equation}\label{SpinN}
S^2=\frac{p_2^2}{m^2}\left[1-\left(\frac{m^2}{M^2}-\frac{{\cal A}_m^2}{M^2}\frac{p_2^2}{m^2}
\right)\right]\,{.}
\end{equation}
Now the second field equation (\ref{stateS}) can be rewritten
in the form \begin{equation}\label{spin2} (\Box_2+m_1^2)(\Box_2+m^2_2)\Phi (q_1, q_2)=0\,{,} \end{equation}
where \begin{equation}\label{Box} \Box_a\equiv \partial_{a\mu}\partial^{a\mu}\,{,} \quad
\partial_{a\mu}= \frac{\partial}{\partial q_a^{\mu}}\,{,}\quad  a=1,2\,{.} \end{equation} There is
no summation with respect to $a$ in Eq.\ (\ref{Box}). The dependence of the quantum
field $\Phi (q_1, q_2)$ on $q_1$ is determined by the Klein--Gordon equation
(\ref{statep1}) with the mass $M$ \begin{equation}\label{qstate} (\Box_1+M^2)\Phi (q_1, q_2)=0\,{.}
\end{equation}
For $M\neq 0$ the parameters $m_a^2,\;\; a=1,2$ in Eq. (\ref{spin2}) are
given by
\begin{equation}
\label{parameter}
 m_{1,2}^2=\frac{m^2}{2}\left[1-\mu^2\pm\sqrt{(1-\mu^2)^2+4S(S+D-1)
\frac{{\cal A}_m^2}{M^2}} \,\right]{,}
 \end{equation}
where $\mu=m/M$. In the massless case $(M=0)$, we
have
\begin{equation}
\label{massless}
m_1^2=0, \quad m_2^2=-\mu_0^2\,m^2\,{.}
 \end{equation}
Thus in general
case in Eq.\ (\ref{spin2}) there are one real `mass' $m_1^2\geq 0$ and one tachyonic
`mass' $m_2^2<0$. The latter is not dangerous here because the physical mass of the
quanta described by the field $\Phi(q_1, q_2)$ is equal to $M$ according to Eq.\
(\ref{qstate}).

From Eqs.\
(\ref{spin2}) and (\ref{qstate})
we deduce the Green function in quantum field theory concerned
 \begin{equation}\label{Green} G(k_1,
k_2)=\frac{1}{(k_1^2-M^2)(k_2^2-m_1^2)(k_2^2-m_2^2)}\,{.} \end{equation} The regularizing
property of this propagator can be elucidated by considering the one--loop Feynman
diagram in such bilocal field theory \begin{equation}\label{Gk1k2} \int\int
G^2(k_1,k_2)d^4k_1d^4k_2\simeq \int\int\frac{d^4k_1}{k_1^4}\frac{d^4k_2}{k_2^8} \sim
\int\frac{d^4k_1}{k_1^4}\int\frac{d^4k_2}{k_2^5(k_1)}\sim \int\frac{d^4k_1}{k_1^4}
\frac{1}{k_1^4}\,{.}
 \end{equation}
 Here we have taken into account that the momenta $k_1$ and
$k_2$ will be, as usually, `mixed' in this diagram describing the interacting field $\Phi
(q_1,q_2)$, and the integration over $dk_2$ has been carried out at first.
As a result, we obtained in Eq. (\ref{Gk1k2}) an additional factor $k_1^{-4}$ in comparison
with the usual one--loop Feynman diagram with the propagator $(k^2-m^2)^{-1}$. This gives
evidence that the maximal acceleration can really provide natural regularization in
corresponding quantum field theory.

\section{Conclusion}

In comparison with other physical regulators of quantum field theory,
for example, elementary length, which also originates in extended nature
of elementary particles, the principle of maximal acceleration is obviously
favorable because it preserves the continuous space--time.

In order to put the arguments in favor of the  regularizing property of the
 maximal acceleration principle onto a more
rigorous footing, it is required to develop the theory of bilocal quantum field
introduced above. A special attention should be paid here on the interaction mechanism
in this formalism. However all these problems are far beyond the scope of this short
note.

\centerline{\bf Acknowledgments}

This work has been accomplished during the stay of one of the authors (V.V.N.) at
Salerno University. It is a pleasant duty for him to thank Prof. G. Scarpetta, Dr. G.
Lambiase, and Dr. A. Feoli for warm hospitality. The work was supported by Russian
Foundation of Fundamental Research (Grant No.\ 97-01-00745) and by fund MURST ex 60\%
and ex 40\% DPR 382/80.

\end{document}